\documentclass[aps,twocolumn,prd,superscriptaddress,noshowpacs,nofootinbib,noshowkeys,floatfix]{revtex4}
\usepackage{graphicx}
\usepackage{amsfonts}
\usepackage{amssymb}
\usepackage{amsmath}

\usepackage{ulem}
\usepackage{color}

\renewcommand\sout{\bgroup \color{blue} \ULdepth=-.5ex \ULset}
\begin{document}

\title{Polyakov loop fluctuations in Dirac eigenmode expansion}

\author{Takahiro~M.~Doi}
\affiliation{%
Department of Physics, Kyoto University,
Kyoto 606-8502,
Japan
}

\author{Krzysztof Redlich}
\affiliation{%
Institute of Theoretical Physics, University of Wroclaw,
PL-50204 Wroclaw,
Poland
}
\affiliation{Extreme Matter Institute EMMI, GSI,
Planckstr. 1, D-64291 Darmstadt, Germany}
\affiliation{Department of Physics, Duke University, Durham, North Carolina 27708, USA}
\author{Chihiro Sasaki}
\affiliation{%
Institute of Theoretical Physics, University of Wroclaw,
PL-50204 Wroclaw,
Poland
}
\affiliation{%
Frankfurt Institute for Advanced Studies,
D-60438 Frankfurt am Main,
Germany
}
\author{Hideo~Suganuma}
\affiliation{%
Department of Physics, Kyoto University,
Kyoto 606-8502,
Japan
}

\begin{abstract}

We investigate correlations of the Polyakov loop fluctuations with 
eigenmodes of the lattice Dirac operator. 
Their analytic relations are derived on the temporally 
odd-number size lattice with the normal non-twisted periodic 
boundary condition for the link-variables. 
We find that the low-lying Dirac modes yield negligible 
contributions to the Polyakov loop fluctuations. 
This property is confirmed to be valid in confined and deconfined phase 
by numerical simulations in SU(3) quenched QCD.
These results indicate that there is no direct, 
one-to-one correspondence between confinement and chiral symmetry breaking 
in QCD in the context of different properties of 
the Polyakov loop fluctuation ratios.

\end{abstract}
\maketitle

\def\slash#1{\not\!#1}
\def\slashb#1{\not\!\!#1}
\def\slashbb#1{\not\!\!\!#1}

\section{Introduction}

Color confinement and chiral symmetry breaking are the striking non-perturbative phenomena
in low-energy QCD, which are of particular importance in particle and nuclear
physics~\cite{NJL, KS, Rothe, Greensite,Fukushima}.

Several scenarios of the confinement mechanism have been proposed~\cite{Rothe, Greensite, NTM_dualSC, Fukushima,GZ, KugoOjima}, where the ghost and gluon propagators in deep infrared need to be
quantified, thus this requires a non-perturbative quantization of QCD.
Whereas this issue has been investigated extensively,
it remains challenging to comprehend the non-perturbative aspects from the
first-principle calculations.

In a pure SU(3) gauge theory, the Polyakov loop is an exact order parameter
of the $Z_3$ center symmetry and for deconfinement, which dictates a first order phase transition~\cite{Rothe,g1,g2,g3}.
In the presence of light dynamical quarks, the  Polyakov loop loses its interpretation
as an order parameter and is smoothly changing  with temperature. However,
contrary to the broad Polyakov loop, a particular ratio of the Polyakov loop
susceptibilities retain a clear remnant of the underlying
$Z_3$ center symmetry
fairly well even in full QCD with the physical pion mass~\cite{LFKRS13q,LFKRS13f}.
Thus,  the ratio of the Polyakov loop fluctuations can serve as
observables to identify the onset of deconfinement in QCD.

In the presence of light dynamical quarks, the transition from hadronic phase
to quark-gluon plasma becomes crossover and accompanies partial restoration of
chiral symmetry  at finite temperature~\cite{ColumbiaPlot,aoki,O(4)}.
Spontaneous chiral symmetry breaking is characterized by a non-vanishing condensation
of quark-bilinear operators.
The low-lying Dirac modes, which are the eigenmodes
of the Dirac operator with small eigenvalues, are known
to be responsible for saturating the chiral condensate of light quarks
$\langle \bar \psi \psi \rangle$,
through the Banks-Casher relation~\cite{BanksCasher}.

In fact, at vanishing and small baryon chemical potential, the lattice QCD results
suggest  that there is an interplay between quark deconfinement and chiral crossovers
as they  take place   in the same narrow temperature range ~\cite{Karsch}.
Also, in the maximally Abelian gauge, 
confinement and chiral symmetry breaking, 
as well as the instantons, simultaneously disappear when the QCD monopoles are
removed~\cite{SNW94, Miyamura, Woloshyn, SM96}.
On the other hand, there exist several observations that unbroken chiral symmetry
does not dictate deconfinement: Given a tower of hadron spectra with
eliminating the low-lying Dirac modes~\cite{LS11}, the hadrons keep
their identity even  in chirally  restored phase where parity
doublers are all degenerate.
In addition, it has been shown in the SU(3) lattice simulations that the low-lying
modes have little contribution to the Polyakov loop and to the confining force,
indicating that the two features are rather independent~\cite{GIS, DSI, SDI}.

In the context of the interplay between confinement and chiral symmetry
breaking~\cite{GIS,Karsch,Miyamura,Woloshyn,aoki,YAoki,LS11,Gattringer,Langfeld,BG08,DSI,SDI},
it is important  to make a reliable separation of one from another, whereas
those phenomena are supposed to be correlated.
The apparent coincidence in change of properties of the Polyakov loop fluctuation ratios
and the chiral condensate and its susceptibility near the  chiral crossover might indicate
that there is a tied relation between confinement  and chiral symmetry breaking in QCD.
However, such a possible relation has  not been  quantified yet.

Utilizing the Dirac-mode expansion method formulated on the lattice~\cite{GIS},
the low-lying modes can be systematically removed in calculating expectation values
of different operators.

In this paper, we apply the above expansion method to
 investigate the relation between confinement and chiral symmetry breaking
in terms of the Polyakov loop fluctuations and their ratios.  We put  our particular
attention to the contribution of  the low-lying Dirac modes
to the Polyakov loop fluctuations.

We derive the  analytic relations 
between the real, imaginary and modulus of the Polyakov loop 
and their fluctuations with  the  Dirac modes   on the temporally 
odd-number  size  lattice with  periodic boundary conditions.
These analytical relations are applicable to both full and quenched 
QCD. 
We show, through numerical simulations on the lattice  in quenched QCD, 
that the low-lying Dirac modes yield negligible contribution to 
the Polyakov loop fluctuations.  
With these results, also not observed is a direct relation 
between confinement and chiral symmetry breaking in QCD 
through the Polyakov loop fluctuation ratios.

The paper is organized as follows:
In the next section we derive a set of analytic relations linking the Polyakov loop
fluctuations to the Dirac eigenmodes.
In Sec. III, we examine the role of the low-lying Dirac modes in the Polyakov loop fluctuations  and
 present our numerical results within quenched lattice QCD.
Section IV is devoted to summary and conclusions.

\section{Polyakov loop fluctuations}

We utilize the SU($N_{\rm c}$) lattice QCD formalism  and
consider a square lattice with spacing $a$.
Each site is indicated by $s=(s_1, s_2, s_3, s_4)$ with  $s_\mu=1,2,\cdots,N_\mu$.
A gauge field, $A_\mu(s) \in SU(N_c)$ with the gauge coupling $g$, is introduced
as a link-variable, $U_\mu(s)={\rm e}^{iagA_\mu(s)}$.
We use spatially symmetric lattice, i.e., $N_1=N_2=N_3\equiv N_\sigma$ and
$N_4\equiv N_\tau$, with $N_\sigma \ge N_\tau$.

\begin{figure}[!t]
\vskip -0.9cm \includegraphics[width=3.60in,angle=-90]{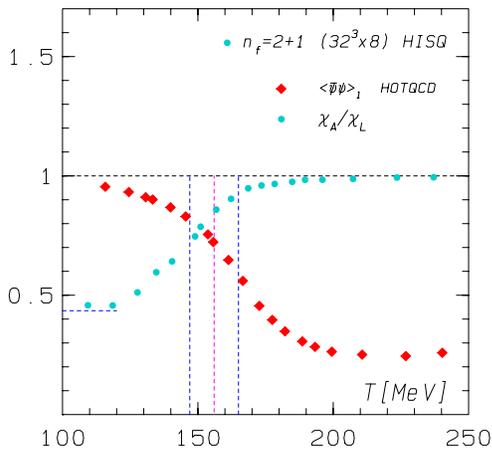}\hskip 0.9cm
 \vskip -1.9cm
\caption{The temperature dependence of the Polyakov loop susceptibilities ratio $R_A=\chi_A/\chi_T$ from Eq. (\ref{RA}) and the chiral condensate of light quarks
$\langle \bar \psi \psi\rangle_l$,  normalized to its zero temperature value. The lattice QCD Monte Carlo results  are from Refs. \cite{LFKRS13f} and \cite{hotqcd}, respectively.  The horizontal dashed lines are the limiting values of $R_A$ in a pure SU(3) lattice gauge theory   \cite{LFKRS13q}. The vertical dashed lines indicate the  chiral crossover temperature and the range of its  errors  \cite{hotqcd,t1,t2}.     }\label{fig1}
 \end{figure}

For each gauge configuration,
the Polyakov loop $L$ is defined as
\begin{align}
L \equiv
\frac{1}{N_{\rm c}V} \sum_s {\rm tr}_c
\{\prod_{i=0}^{N_\tau-1}U_4(s+i\hat{4})\}, \label{Polyakov}
\end{align}
where $\hat{\mu}$ is the unit vector in direction of $\mu$ in the lattice unit,
and $V$ is the 4-dimensional lattice volume, $V=N_\sigma^3N_\tau$.

Under the $Z_3$ rotation, the Polyakov loop is transformed into
\begin{align}
\tilde{L}
=
L \mathrm{e}^{2\pi k i/3} \label{TransformedPolyakov}
\end{align}
with $k=0,\pm1$~\cite{LFKRS13q,LFKRS13f}.
In confined phase, $k=0$ is taken.
In deconfined phase, $k$ is chosen such that
the transformed Polyakov loop $\tilde{L}$ lies in its real sector.

We introduce the Polyakov loop susceptibilities,
\begin{align}
T^3\chi_A=\frac{N_\sigma^3}{N_\tau^3}[\langle |L|^2 \rangle-\langle |L| \rangle^2], \label{ChiA}\\
T^3\chi_L=\frac{N_\sigma^3}{N_\tau^3}[\langle (L_L)^2 \rangle-\langle L_L \rangle^2], \label{ChiL}\\
T^3\chi_T=\frac{N_\sigma^3}{N_\tau^3}[\langle (L_T)^2 \rangle-\langle L_T \rangle^2], \label{ChiT}
\end{align}
where $L_L \equiv {\rm Re}(\tilde{L})$ and $L_T \equiv {\rm Im}(\tilde{L})$,
and consider their ratios,
\begin{align}
R_A \equiv \frac{\chi_A}{\chi_L}, \label{RA}\\
R_T \equiv \frac{\chi_T}{\chi_L}. \label{RT}
\end{align}
The Polyakov loop  susceptibility ratios (\ref{RA}) and (\ref{RT}) were shown to be
excellent probes of the deconfinement phase transition in a pure gauge theory~\cite{LFKRS13q,LFKRS13f}.
They are almost temperature independent above and below the transition and exhibit
a discontinuity at the transition temperature. This characteristic behavior is
understood in terms of the global $Z_3$  symmetry of the Yang-Mills
Lagrangian and the general properties of the Polyakov loop probability distribution~\cite{LFKRS13q}.

In the presence of dynamical quarks, the Polyakov loop is no longer an order
parameter and stays finite even in the low temperature phase.  Consequently,
the ratios of the Polyakov loop susceptibilities are modified due to  explicit
breaking of the $Z_3$ center symmetry.  Indeed, both  $R_A$ and $R_T$
vary continuously with temperature across the chiral crossover, however  $R_A$
interpolates between the two limiting values set by the pure gauge theory.
This property of $R_A$  is illustrated in  Fig. \ref{fig1}.
Also seen in this figure is that,
in spite of  smoothening effects observed in the presence of quarks, there is
an  abrupt rate change with $T$ in $R_A$   near the chiral crossover $T \simeq 155$ 
MeV~\cite{LFKRS13f}. 
The ratio $R_A$ in Fig. \ref{fig1} has its inflection point at 
$T\simeq 150 $ MeV, which is fairly in good agreement 
with the chiral crossover range calculated in 
lattice QCD. 
Such behavior of the Polyakov loop fluctuation  ratio can be
regarded as  an effective  observable indicating deconfinement properties
in QCD \cite{LFKRS13f}.

The apparent  modification of $R_A$ and $R_T$   near chiral crossover may
suggest that there are certain correlations between confinement  and chiral
symmetry breaking. Such correlations can be best verified  when expanding the
Polyakov loop and its fluctuations in terms of the Dirac eigenmodes.

In the
following, we formulate the relevant quantities,  based on this expansion method,
and study the influence of the low-lying Dirac modes on the properties of the
Polyakov loop fluctuation ratios.

\section{Dirac-mode expansion} 

{}To derive  the analytic relation between the
Polyakov loop and the Dirac modes, we consider  the temporally odd-number
lattice with the normal non-twisted periodic boundary condition for link-variables,
in both temporal and spatial directions~\cite{SDI, DSI}.


We introduce the link-variable operator $\hat{U}_{\pm\mu}$,
with the matrix element
\begin{align}
\langle
s | \hat{U}_{\pm\mu} |s' \rangle=U_{\pm\mu}(s)\delta_{s\pm\hat{\mu},s'}. \label{LinkOp}
\end{align}

The covariant derivative operator on the lattice is introduced as
\begin{align}
\hat{D}_\mu= \frac{1}{2a}(\hat{U}_{\mu}-\hat{U}_{-\mu}),   \label{CovariantOp}
\end{align}
and the Dirac operator
\begin{align}
\hat{\slashb{D}}=\gamma_\mu \hat{D}_\mu=
\frac{1}{2a}\sum_{\mu=1}^{4}\gamma_\mu (\hat{U}_{\mu}-\hat{U}_{-\mu}),   \label{DiracOp}
\end{align}
with  its  matrix element  
\begin{align}
 \slashb{D}_{s,s'}
      = \frac{1}{2a} \sum_{\mu=1}^4 \gamma_\mu
\left[ U_\mu(s) \delta_{s+\hat{\mu},s'}
        - U_{-\mu}(s) \delta_{s-\hat{\mu},s'} \right], \label{DiracOpExp}
\end{align}
where $U_{-\mu}(s)= U^\dagger_\mu(s-\hat\mu)$ and
$\gamma^\dagger_\mu=\gamma_\mu$.

Since the Dirac operator is anti-hermitian,
the Dirac eigenvalue equation reads 
\begin{eqnarray}
\hat{\slashb{D}}|n\rangle =i\lambda_n|n \rangle,
\end{eqnarray}
where $\lambda_n \in {\bf R}$.
Using the Dirac eigenfunction $\psi_n(s) \equiv \langle s|n \rangle $,
one arrives at the eigenvalue equation
\begin{eqnarray}
 \frac{1}{2a}& \sum_{\mu=1}^4 \gamma_\mu
[U_\mu(s)\psi_n(s+\hat \mu)-U_{-\mu}(s)\psi_n(s-\hat \mu)] \ \ \ \ \ & \nonumber \\
& \ \ \ \ \ \ \ \ \ \ \ \ \ \ \ \ \ \ \ \ \ \ \ \ \ \ \ \ \ \ \ \ \ \ \ \ \ \ \ \ \ \ \ =i\lambda_n \psi_n(s).
\label{DiracEigenExp}
\end{eqnarray}

At finite temperature, 
imposing the temporal anti-periodicity 
for $\hat D_4$ acting on quarks, it is convenient 
to add a minus sign to the matrix element of 
the temporal link-variable operator $\hat U_{\pm 4}$ 
at the temporal boundary of $t=N_t(=0)$ \cite{SDI}:
\begin{eqnarray}
\langle {\bf s}, N_t|\hat U_4| {\bf s}, 1 \rangle 
&=&-U_4({\bf s}, N_t),
\nonumber \\
\langle {\bf s}, 1|\hat U_{-4}| {\bf s}, N_t \rangle 
&=&-U_{-4}({\bf s}, 1)=-U_4^\dagger({\bf s}, N_t). 
\label{eq:LVthermal}
\end{eqnarray}
Then, the Polyakov loop in Eq. (\ref{Polyakov}) is expressed as
\begin{align}
L
=-\frac{1}{N_{\rm c}V}{\rm Tr}_c \{\hat U_4^{N_\tau}\} 
=\frac{1}{N_c V}
\sum_s {\rm tr}_c \{\prod_{n=0}^{N_t-1} U_4(s+n\hat t)\},
\label{PolyakovOp}
\end{align}
where Tr$_c$ denotes the functional trace,
${\rm Tr}_c \equiv \sum_s {\rm tr}_c$,
and ${\rm tr}_c$ is taken over color index.
The minus sign stems from the additional minus on $U_4({\bf s}, N_t)$ 
in Eq.(\ref{eq:LVthermal}).

Note that the functional trace of a product of link-variable operators
corresponding to non-closed path is exactly zero. Indeed, followed by
Eq.(\ref{LinkOp}), one obtains
\begin{align}
&{\rm Tr}_c(\hat{U}_{\mu_1}\hat{U}_{\mu_2}\cdots\hat{U}_{\mu_{N_P}})
=
{\rm tr}_c\sum_s\langle s|\hat{U}_{\mu_1}\hat{U}_{\mu_2}\cdots\hat{U}_{\mu_{N_P}}|s\rangle \nonumber \\
=&{\rm tr}_c\sum_s U_{\mu_1}(s)\cdots U_{\mu_{N_P}}(s+\sum_{k=1}^{N_P-1}\hat{\mu}_k)
\langle s+\sum_{k=1}^{N_P}\hat{\mu}_k|s\rangle \nonumber \\
=&0,
\label{nonclosed}
\end{align}
with $\sum_{k=1}^{N_P}\hat{\mu}_k\neq 0$ for any non-closed path
of length $N_P$.
This is   understood from Elitzur's theorem~\cite{Elitzur},  that
the vacuum expectation values of gauge-variant operators are zero.

In the following, we show that the Polyakov loop can be explicitly expanded
in terms  of eigenmodes of the Dirac operator.

\subsection{Relation between the Polyakov loop and Dirac modes}

\begin{figure}
\begin{center}
\includegraphics[scale=0.4]{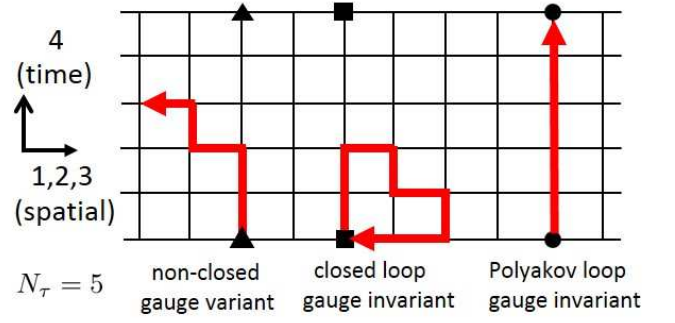}
\caption{ The link path structure on a  temporally odd-number lattice with  $N_\tau=5$ and with periodic boundary condition.
The left configuration is gauge-variant,
whereas the middle is gauge-invariant.
The right configuration represents a  closed path  with $N_\tau$-(link-variables) because of the periodicity in a temporal direction,
thus this path  corresponds to a gauge-invariant Polyakov loop.
}
\label{OddLattice}
\end{center}
\end{figure}
We introduce the following key quantity,
\begin{eqnarray}
I={\rm Tr}_{c,\gamma} (\hat{U}_4\hat{\slashb{D}}^{N_\tau-1}),  \label{I}
\end{eqnarray}
where
${\rm Tr}_{c,\gamma}\equiv \sum_s {\rm tr}_c
{\rm tr}_\gamma$, and  ${\rm tr}_\gamma$ is taken over spinor indexes.
From the definition of  $\hat{\slashb{D}}$   in Eq.(\ref{DiracOp}),
it is clear that $\hat{U}_4\hat{\slashb{D}}^{N_\tau-1}$ operator is expressed
by a sum of products  of $N_\tau$- (link-variables).

Note that, on a square lattice, it is not possible to construct
any closed loop using the products of odd number of link-variables
(see Fig. \ref{OddLattice} for illustration).
By construction, one considers a square lattice with odd $N_\tau$, thus
$\hat{U}_4\hat{\slashb{D}}^{N_\tau-1}$  in Eq. (\ref{I}) does not contain any
contributions from the products of $N_\tau$-(link-variables) along the closed paths.
However, due to the periodic boundary condition in time direction, the only
exception is a gauge-invariant term  proportional to $\hat{U}_4^{N_\tau}$.
This term is thus related  with  the Polyakov loop operator.

Based on the above discussion, and applying  Eqs. (\ref{PolyakovOp}),
(\ref{nonclosed}) and (\ref{CovariantOp}) to Eq. (\ref{I}), one finds that
\begin{align}
I
=\frac{12V}{(2a)^{N_\tau-1}}L\,, \label{I1}
\end{align}
thus $I$  is directly proportional to the Polyakov loop.

On the other hand,
since $I$ in Eq. (\ref{I}) is defined through the functional trace,
it can be expressed in the basis of Dirac eigenmodes as
\begin{align}
I
&=\sum_n\langle n|\hat{U}_4\slashb{\hat{D}}^{N_\tau-1}|n\rangle \nonumber\\
&=i^{N_\tau-1}\sum_n\lambda_n^{N_\tau-1}\langle n|\hat{U}_4| n \rangle.  \label{I2}
\end{align}

Consequently, from  Eqs.(\ref{I1}) and  (\ref{I2}) one finds
that, on the temporally odd-number lattice,  there is a direct relation between
the Polyakov loop and the Dirac modes~\cite{SDI, DSI},
\begin{eqnarray}
L=\frac{(2ai)^{N_\tau-1}}{12V}
\sum_n\lambda_n^{N_\tau-1}\langle n|\hat{U}_4| n \rangle
,  \label{RelOrig}
\end{eqnarray}
which is a Dirac spectral representation of the Polyakov loop.

The relation (\ref{RelOrig}) is a mathematical identity, 
and is exactly satisfied for arbitrary gauge configurations. 
Consequently, this relation is valid, regardless of whether 
the link-variables are generated in full QCD or quenched QCD \cite{SDI}. 

The relation (\ref{RelOrig}) enables to investigate the contribution of different
Dirac-modes to the Polyakov loop. Of particular interest is the role of the low-lying
eigenvalues which are essential to identify chiral symmetry restoration in QCD
at finite temperature.

In principle, it is possible to calculate numerically the contribution of the
Dirac modes  to the Polyakov loop through Eq. (\ref{RelOrig}) by  using   Monte
Carlo simulations. However,  a very  large $(4\times N_c\times V)^2$--dimension
of a Dirac operator implies also  a large cost of  numerical calculations.
This can be, however,  remarkably reduced by use of
the modified Kogut-Susskind(KS) formalism \cite{DSI}.

We rewrite Eq.~(\ref{RelOrig}) into the equivalent form
\begin{align}
 L =\frac{(2ai)^{N_\tau-1}}{3V}
\sum_{n}\lambda_n^{N_\tau-1}(n|\hat{U}_4|n), \label{RelKS}
\end{align}
where the KS Dirac eigenstate $|n)$ is obtained by solving the eigenvalue
equation,
\begin{align}
\eta_\mu D_\mu|n) =i\lambda_n|n ). \label{KSEigenEq}
\end{align}
Here, the KS Dirac operator $\eta_\mu D_\mu$ is defined as
\begin{align}
(\eta_\mu D_\mu)_{ss'}=\frac{1}{2a}\sum_{\mu=1}^{4}\eta_\mu(s)\left[U_\mu(s)\delta_{s+\hat{\mu},s'}-U_{-\mu}(s)\delta_{s-\hat{\mu},s'}\right] \label{KSDiracOp}
\end{align}
with the staggered phase $\eta_\mu(s)$,
\begin{align}
\eta_1(s)\equiv 1, \ \ \eta_\mu(s)\equiv (-1)^{s_1+\cdots+s_{\mu-1}}
\ (\mu \geq 2). \label{eta}
\end{align}
Consequently, each Dirac-mode contribution to the Polyakov loop is obtained
by solving the eigenvalue equation of the KS Dirac operator whose dimension is now
$(N_c\times V)^2$,
instead of $(4\times N_c\times V)^2$, as in the original Dirac operator.

In terms of the KS Dirac eigenfunction $\chi_n(s)= \langle s|n)$,
the KS Dirac matrix element $(n|\hat{U}_\mu|m)$ is explicitly expressed as
\begin{align}
(n|\hat{U}_\mu|m)
&=\sum_s (n|s \rangle \langle s|\hat{U}_\mu| s+\hat{\mu} \rangle\langle s+\hat{\mu}|m ) \nonumber\\
&=\sum_s \chi_n(s)^\dagger U_\mu(s) \chi_m(s+\hat{\mu}).
\end{align}
{}From the gauge transformation property of link-variables and the KS Dirac eigenfunctions,
the matrix element $(n|\hat{U}_\mu|m)$ is gauge-invariant \cite{SDI,DSI}.

Note that the modified KS formalism applied here is not an approximation,
but is a method for spin-diagonalization of the Dirac operator.
In this study, we do not use specific fermions, such as the KS fermion,
but apply  the modified KS formalism as a prescription to reduce the numerical cost.

The  relations  of the Polyakov loop and  Dirac eigenmodes in Eqs.  (\ref{RelOrig})
and  (\ref{RelKS}) are exact. They are valid at finite temperature and density,
and are independent of the particular implementation of fermions on the lattice~\cite{SDI},  thus can be used to identify the interplay between deconfinement and   chiral
symmetry restoration in QCD.

\subsection{The  Polyakov loop fluctuations and Dirac modes}

The expansion of the Polyakov loop in terms of the Dirac eigenmodes, formulated
in the previous section, can be also applied to the fluctuations of the real,
imaginary and  modulus of the Polyakov loop.

Multiplying Eq. (\ref{RelOrig}) by the $Z_3$ factor $\mathrm{e}^{2\pi ki/3}$,
one obtains the relation between the $Z_3$ transformed Polyakov loop $\tilde{L}$ and
the Dirac modes,
\begin{eqnarray}
\tilde{L}=\frac{(2ai)^{N_\tau-1}}{12V}
\sum_n\lambda_n^{N_\tau-1}\mathrm{e}^{2\pi ki/3}\langle n|\hat{U}_4| n \rangle,
  \label{LtildeDirac}
\end{eqnarray}
where $k=0, \pm1$ is chosen such that, for each gauge configuration,
the $\tilde{L}$ lies in a  real sector.

Taking the real and the imaginary parts of Eq. (\ref{LtildeDirac}), the Dirac
spectral representation of the longitudinal and transverse Polyakov loops reads
\begin{eqnarray}
L_L=\frac{(2ai)^{N_\tau-1}}{12V}
\sum_n\lambda_n^{N_\tau-1}{\rm Re}\left(\mathrm{e}^{2\pi ki/3}\langle n|\hat{U}_4| n \rangle\right),
\label{LLDirac} \\
L_T=\frac{(2ai)^{N_\tau-1}}{12V}
\sum_n\lambda_n^{N_\tau-1}{\rm Im}\left(\mathrm{e}^{2\pi ki/3}\langle n|\hat{U}_4| n \rangle \right),
\label{LTDirac}
\end{eqnarray}
respectively, whereas,  taking  the absolute value of Eq. (\ref{RelOrig}),
the following relation is also obtained;
\begin{eqnarray}
|L|=\frac{(2a)^{N_\tau-1}}{12V}
\left|\sum_n\lambda_n^{N_\tau-1}\langle n|\hat{U}_4| n \rangle\right|. \label{LabsDirac}
\end{eqnarray}

Since Eqs. (\ref{LtildeDirac}), (\ref{LLDirac}), (\ref{LTDirac})  and (\ref{LabsDirac}) are valid
for each gauge configuration, the Dirac spectral representation for different
fluctuations of  the Polyakov loop and their ratios are directly obtained by
substituting Eqs. (\ref{LLDirac})-(\ref{LabsDirac}) to
Eqs. (\ref{ChiA})-(\ref{ChiT}).
As an example we quote an explicit  expression for the Dirac spectral representation
of the $R_A=\chi_A/\chi_L$ ratio as
\begin{widetext}
\begin{align}
R_A=
\frac{
\left\langle\left|\sum_n\lambda_n^{N_\tau-1}\langle n|\hat{U}_4| n \rangle\right|^2\right\rangle
-
\left\langle\left|\sum_n\lambda_n^{N_\tau-1}\langle n|\hat{U}_4| n \rangle\right|\right\rangle^2
}{
\left\langle\left(\sum_n\lambda_n^{N_\tau-1}{\rm Re}\left(\mathrm{e}^{2\pi ki/3}\langle n|\hat{U}_4| n \rangle\right)\right)^2\right\rangle
-
\left\langle\sum_n\lambda_n^{N_\tau-1}{\rm Re}\left(\mathrm{e}^{2\pi ki/3}\langle n|\hat{U}_4| n \rangle\right)\right\rangle^2
},
\end{align}
\end{widetext}
where $\langle x \rangle$ denotes an average over all gauge configurations.

The explicit analytic relations of the Dirac spectral decomposition 
of the real,imaginary and modulus Polyakov loop and their fluctuations are 
the key results of our studies. 
Note here that, like Eq.(\ref{RelOrig}), 
these relations (\ref{LtildeDirac})-(\ref{LabsDirac}) 
are applicable to both full and quenched QCD, 
since we just use Elitzur's theorem~\cite{Elitzur}: 
only gauge-invariant quantities survive.
All these relations are derived on the temporally odd-number lattice 
for practical reasons. 
However, a particular choice of the parity for the lattice size in time direction 
does not alter the physics,  since in the continuum limit, $a\rightarrow 0$ and 
$N_\tau\rightarrow \infty$, any number of large $N_\tau$ should give the same result \cite{DSI,SDI}. 
In fact, similar relations are derived also on the even lattice, 
whereas a more compact form can be obtained 
on the temporally odd-number lattice \cite{SDI,DSI}. 
It is however difficult to take the continuum extrapolation.
For instance, the continuum limit of the Polyakov loop itself is still
unsettled because of uncertainty of its renormalization. 
However, at least, the ambiguity of the multiplicative renormalization of 
the Polyakov loop can be avoided by considering the ratio of 
the Polyakov-loop susceptibilities \cite{LFKRS13q,LFKRS13f}.

\section{Numerical  results}

{}To study numerically the influence of different Dirac modes on the Polyakov loop
fluctuations and their ratios,
we further apply the modified KS formalism. This amounts in replacing
the diagonal Dirac matrix element $\langle n|\hat{U}_4| n \rangle$ in
Eqs. (\ref{LLDirac})-(\ref{LabsDirac}) by the corresponding
KS Dirac matrix element $( n|\hat{U}_4| n )$ \cite{DSI}, as
\begin{align}
\langle n|\hat{U}_4| n \rangle = 4( n|\hat{U}_4| n ).
\end{align}

We analyze the contributions from 
the low-lying Dirac modes to the Polyakov loop fluctuations 
in the SU(3) lattice QCD through  Monte Carlo simulations. 
In the mathematical sense, all the obtained relations 
(\ref{LtildeDirac})-(\ref{LabsDirac}) 
hold for both full and quenched QCD. 
In this paper, we perform SU(3) lattice QCD calculations 
with the standard plaquette action 
at the quenched level on $10^3\times5$ size lattice.
Numerical studies are carried out both in confined  and deconfined phases for
different couplings  $\beta=\frac{2N_{\rm c}}{g^2}$, and the corresponding
temperatures,  $T=1/(N_\tau a)$. 
We use the Linear Algebra PACKage (LAPACK) \cite{LAPACK}
in diagonalizing the KS Dirac operator to obtain 
the eigenvalues $\lambda_n$ and the eigenfunctions $\chi_n(s)$. 
The lattice spacing $a$ is 
determined by the zero-temperature string tension of 
$\sigma =0.89$ GeV/fm on a large lattice at each $\beta=6/g^2$. 
In fact, we here calculate the static quark-antiquark potential $V(r)$ 
on 
$16^4$ lattice
 at each $\beta$, and fit it by the Cornell potential, 
i.e., the Coulomb plus linear form \cite{Rothe}, 
to extract the string tension $\sigma$. 

In confined phase, we fix $\beta= 5.6$ on 
$10^3\times5$ lattice, 
which corresponds to  $a\simeq0.25$ fm and  $T\simeq160$ MeV. 
We also calculate the Creutz ratio $\chi(3,3) \simeq 0.35(2)$, 
for an estimate of the string tension or the lattice spacing 
on the precise lattice \cite{Rothe,CM82}, 
in spite of an additional contamination from the Coulomb potential. 
Nevertheless, 
this value is consistent with the string tension $\sigma$ 
obtained from the potential $V(r)$, and leads to $a \simeq 0.28$ fm. 
Here, the average plaquette value is obtained as
$\langle U_{\mu\nu}\rangle$=0.53(2), 
which is consistent
with 
the previous SU(3) lattice studies \cite{CM82}. 
In deconfined phase, the simulations are performed at $\beta = 6.0$ 
on $10^3 \times 5$ lattice, i.e., for $a\simeq0.1$ fm and  $T\simeq400$ MeV. 
On this lattice, the average plaquette value is 
$\langle U_{\mu\nu}\rangle$=0.60(2), 
which is also consistent
with 
the previous works \cite{CM82}.
For each value of $\beta$, we use 20 gauge configurations,
which are taken every 500 sweeps after the thermalization of 5000 sweeps.

Since the  Polyakov loop and its different fluctuations
are expressed as the sums over all  Dirac-modes,
we divide the entire tower of the Dirac eigenvalues  into the low- and higher-lying
modes with the insertion of the  infrared cutoff $\Lambda$.

Based on Eq.(\ref{RelKS}),
we introduce the $\Lambda$--dependent  Polyakov loops,
\begin{align}
&|L|_{\Lambda}=\frac{(2a)^{N_\tau-1}}{3V}
\left|\sum_{|\lambda_n|>\Lambda}\lambda_n^{N_\tau-1}( n|\hat{U}_4| n )\right|, \label{LabsDiraccutKS}
\end{align}
for the modulus, and
\begin{align}
&(L_L)_{\Lambda}=C_{\tau}
\sum_{|\lambda_n|>\Lambda}
\lambda_n^{N_\tau-1}{\rm Re}\left(\mathrm{e}^{2\pi ki/3}( n|\hat{U}_4| n )\right),
\label{LLDiraccutKS} \\
&(L_T)_{\Lambda}=C_{\tau}
\sum_{|\lambda_n|>\Lambda}
\lambda_n^{N_\tau-1}{\rm Im}\left(\mathrm{e}^{2\pi ki/3}( n|\hat{U}_4| n ) \right).
\label{LTDiraccutKS}
\end{align}
for the real and the imaginary part, respectively,
with $C_{\tau}=(2ai)^{N_\tau-1}/{3V}$.

Applying the cutoff dependent Polyakov loops from Eqs. (\ref{LabsDiraccutKS}),
(\ref{LLDiraccutKS}) and (\ref{LTDiraccutKS}) to Eqs. (\ref{ChiA})-(\ref{ChiT}),
we also introduce the $\Lambda$--dependent susceptibilities
\begin{align}
&T^3(\chi)_{\Lambda}=\frac{N_\sigma^3}{N_\tau^3}
[\langle Y_{\Lambda}^2 \rangle-\langle Y_{\Lambda} \rangle^2], \label{ChiAcut}
\end{align}
where $Y$ stands for  $|L|$, $L_L$ or $L_T$, and their ratios
\begin{align}
&(R_A)_{\Lambda}=\frac{(\chi_A)_{\Lambda}}{(\chi_L)_{\Lambda}},~~
&(R_T)_{\Lambda}=\frac{(\chi_T)_{\Lambda}}{(\chi_L)_{\Lambda}} \label{RTcut}.
\end{align}

{}To differentiate the importance of the low-lying Dirac modes on the properties
of the Polyakov loop fluctuations and the chiral condensate, we introduce the
cutoff-dependent chiral condensate $\langle \bar{\psi}\psi\rangle_{\Lambda}$.
In terms of the Dirac modes,  the $\langle \bar{\psi}\psi\rangle_{\Lambda}$  is
expressed as \cite{GIS}
\begin{align}
\langle \bar{\psi}\psi \rangle_{\Lambda}
=-\frac{1}{V}\sum_{|\lambda_n|\geq\Lambda}\frac{2m}{\lambda_n^2+m^2},
\end{align}
where $m$ is the current quark mass.

The chiral condensate is strongly affected by the low-lying Dirac modes.
Taking a  typical value for the infrared  cutoff $\Lambda\simeq0.4\  {\rm GeV}$  and
the quark mass $m\simeq5 \ {\rm MeV}$,  leads to a drastic reduction
of the chiral condensate
\begin{align}
R_{\rm chiral}= \frac{\langle \bar{\psi}\psi\rangle_{\Lambda}}
{\langle \bar{\psi}\psi\rangle}\simeq 0.02, \label{R_chiral}
\end{align}
in a confined phase at $T \simeq 0$ ~\cite{GIS,DSI}.

\begin{figure}[h]
\begin{center}
\includegraphics[scale=0.5]{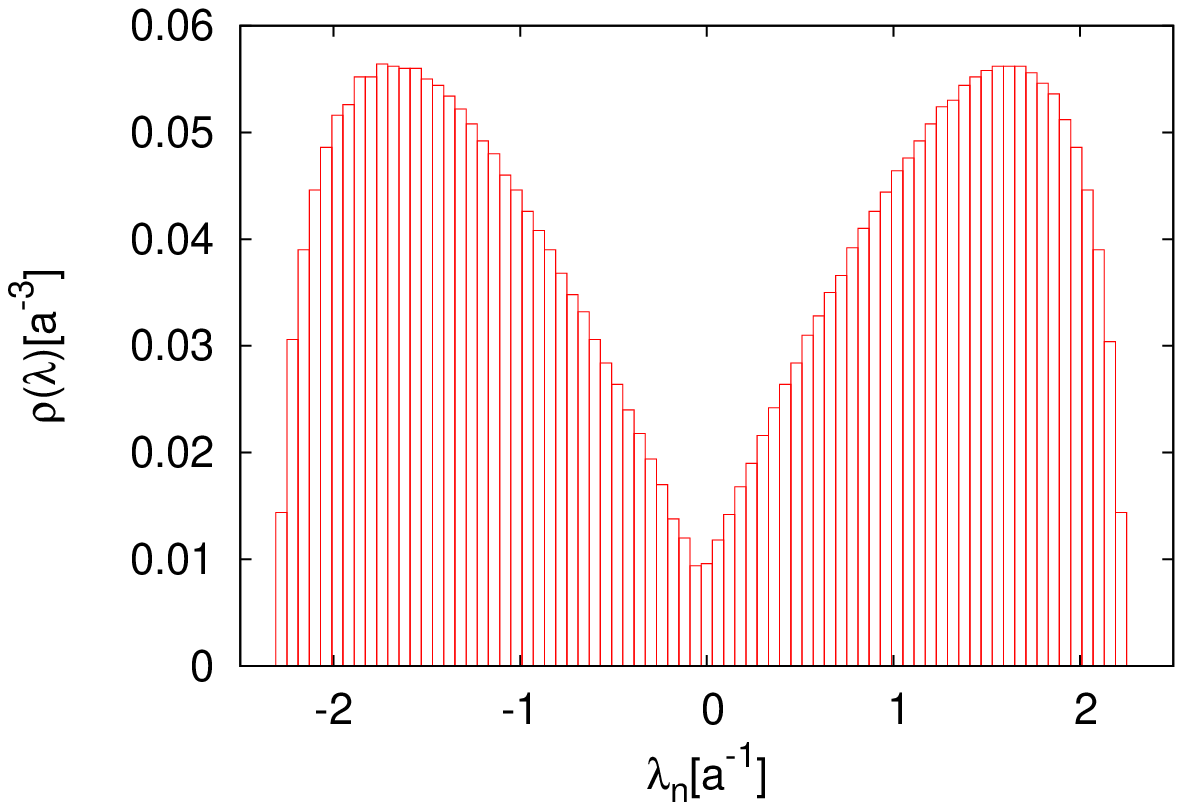}
\includegraphics[scale=0.5]{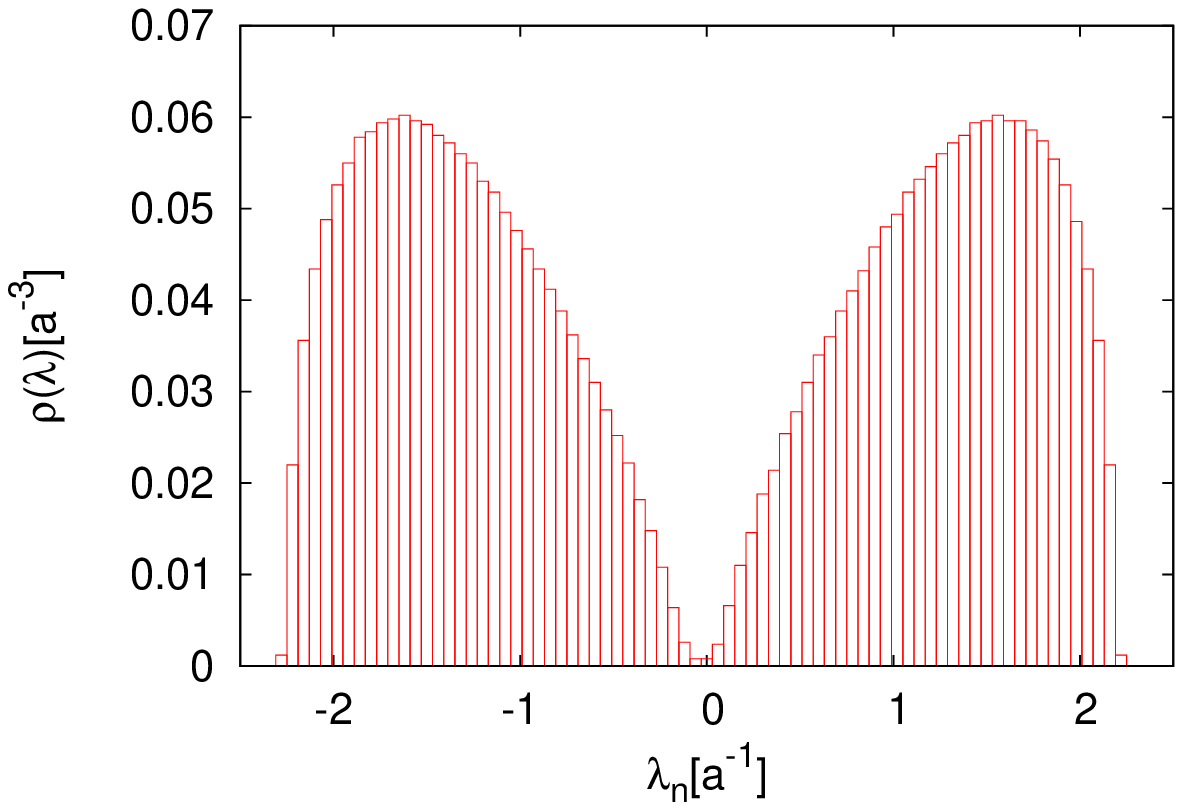}
\caption{
The lattice QCD result of the Dirac eigenvalue distribution 
$\rho(\lambda)$ in the lattice unit:
(a) the confinement phase 
at $\beta = 5.6$ (i.e., $a\simeq0.25$ fm) on $10^3\times5$; 
(b) the deconfinement phase 
at $\beta = 6.0$ (i.e., $a\simeq0.10$ fm) on $10^3\times5$ \cite{DSI}. 
}
\label{DiracEigenDist}
\end{center}
\end{figure}

\begin{figure}
\begin{center}
\includegraphics[width=8cm]{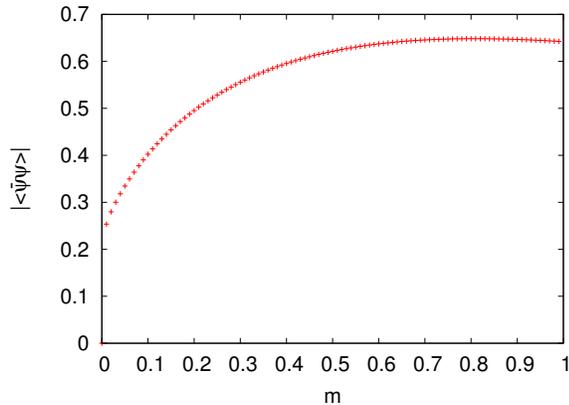}
\caption{
The absolute value of bare chiral condensate 
$|\langle \bar{\psi}\psi \rangle|$ per a flavor 
in the confinement phase 
plotted against the quark mass $m$ in the lattice unit.
The lattice QCD calculation is done at 
$\beta=5.6$ (i.e., $a\simeq0.25$ fm) on $10^3\times 5$.}
\label{Mass_vs_qbarq}
\end{center}
\end{figure}

Figure~\ref{DiracEigenDist} shows the Dirac eigenvalue distribution 
$\rho(\lambda)=1/V\sum_n\langle\delta(\lambda-\lambda_n)\rangle$ 
in confined ($\beta=5.6$) and deconfined ($\beta=6.0$) phases. 
Note here that the near-zero Dirac-mode density 
$\rho(\lambda \simeq 0)$ is apparently finite in confined phase, 
whereas it is highly suppressed in deconfined phase.
We show in Fig.~\ref{Mass_vs_qbarq}  
the bare chiral condensate $|\langle \bar{\psi}\psi \rangle|$ 
per a flavor in the confined phase 
as a function of the quark mass $m$ in the lattice unit. 
The chiral condensate remains finite 
in the small-$m$ region. 
Thus, from Figs.~\ref{DiracEigenDist} and \ref{Mass_vs_qbarq}
it is clear that 
the chiral symmetry is definitely broken in confined phase, 
whereas it is restored in deconfined phase. 

In the same spirit, we introduce the  ratio,
\begin{align}
R_{\rm conf}= \frac{(R_A)_{\Lambda}}{R_A}\,, \label{R_conf}
\end{align}
to quantify the  sensitivity of  the Polyakov loop fluctuations to the particular
Dirac modes.
When, with some  $\Lambda$,  the ratio stays $R_{\rm conf}\simeq 1$,
then  the low-lying Dirac modes below the cutoff $\Lambda$
have a negligible contribution to the Polyakov loop fluctuations.

\begin{figure}
\begin{center}
\includegraphics[width=8cm]{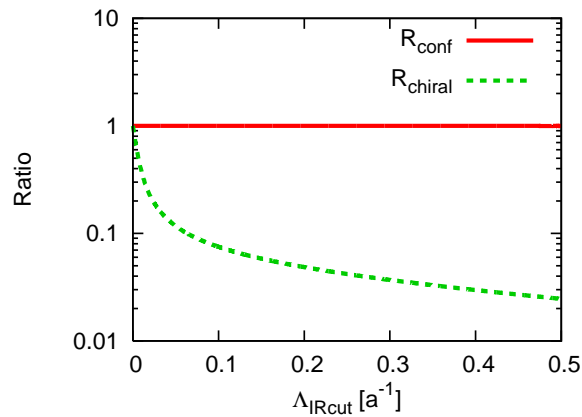}
\caption{
The numerical results for the  $R_{\rm chiral}$ and $R_{\rm conf}$ ratio  from
Eqs. (\ref{R_chiral}) and (\ref{R_conf}), respectively,
as a function of
an  infrared  cutoff $\Lambda$ introduced on   Dirac eigenvalues, expressed
in lattice units.
The Monte Carlo calculations have  been performed
on   $10^3\times 5$  lattice at  $\beta=5.6$, and for
the quark mass of $m=5 \ {\rm MeV}$.
}
\label{ratio_conf}
\end{center}
\end{figure}

In Fig.~\ref{ratio_conf}, we show the Monte Carlo results for  $R_{\rm conf}$
in a confined phase at $\beta=5.6$, for various values of the infrared  cutoff $\Lambda$.
For the sake of comparison, we also show in  Fig.~\ref{ratio_conf} the
$R_{\rm chiral}$ ratio, calculated at the same temperature and with the light
quark mass,  $m=5 \ {\rm MeV}$.
The ratios, $R_{\rm conf}$ and $R_{\rm chiral}$, indicate
the influence of removing the low-lying Dirac modes with the infrared cutoff $\Lambda$
on confinement and chiral symmetry breaking, respectively.

{}From Fig. \ref{ratio_conf}, it is clear that
the $R_{\rm chiral}$ ratio is strongly reduced
by removing the low-lying Dirac modes.
Thus, the low-lying Dirac modes, which are important modes for chiral symmetry breaking, 
are also dominant  to  quantify the chiral condensate.
In contrast to $R_{\rm chiral}$, 
the $R_{\rm conf}$ ratio is almost unchanged when  removing the low-lying Dirac modes
even with relatively large cutoff $\Lambda\simeq 0.5 \ {\rm GeV}$.

Thus, the essential modes for chiral symmetry breaking in QCD
are not important to quantify the Polyakov loop fluctuation ratios, which are
sensitive observables to confinement properties  in QCD.
The same result is also found in deconfined phase, as seen in Table~\ref{ConfIRcut},
which summarizes our numerical results on different fluctuations of the Polyakov loop
and their ratios, obtained on the lattice at $\beta=5.6$ and $\beta=6.0$,
with 20 gauge configurations.
Note here that 
the analytical relation (\ref{RelOrig}) 
and the subsequent formulae of Eqs.(\ref{LLDirac})-(\ref{LabsDirac}) 
hold for each gauge configuration, and 
the contribution from the low-lying Dirac modes 
to the Polyakov loop $L$ is found to be negligible \cite{DSI}.
This fact inevitably leads to almost equivalence between 
the Polyakov-loop fluctuations and those without the low-lying Dirac modes, 
although the statistical error is 
significant because of the small statistics. 

\begin{table}
\caption{
Numerical results for different Polyakov-loop fluctuations (original),
and those without the low-lying Dirac modes (IR-removed)
below the IR-cutoff $\Lambda\simeq 0.4 \ {\rm GeV}$.
The results are obtained in quenched  QCD on
 $10^3\times5$ lattice at  $\beta=5.6$ (confined phase)
and $\beta=6.0$ (deconfined phase) 
with 20 gauge configurations.
}
\begin{tabular}{cccc} \hline \hline
$\beta$ & & original & IR-removed \\ \hline
$\beta$=5.6 & ~~~~$T^3\chi_A$~~~~
&$3.475\times10^{-4}$ &$3.470\times10^{-4}$\\
& $T^3\chi_L$
&$5.307\times10^{-4}$ &$5.298\times10^{-4}$\\
& $T^3\chi_T$
&$6.005\times10^{-4}$ &$5.994\times10^{-4}$\\
& $R_A$
&0.6548 &0.6549 \\
& $R_T$
&1.131 &1.131 \\
\\
$\beta$=6.0 & $T^3\chi_A$
&$2.965\times10^{-3}$ &$2.965\times10^{-3}$\\
& $T^3\chi_L$
&$3.015\times10^{-3}$ &$3.015\times10^{-3}$\\
& $T^3\chi_T$
&$7.848\times10^{-4}$ &$7.848\times10^{-4}$\\
& $R_A$
&0.9834 & 0.9834 \\
& $R_T$
&0.2603 & 0.2603 \\ \hline \hline
  \end{tabular}
\label{ConfIRcut}
\end{table}

The differences in the influence of the low-lying Dirac modes  on the chiral
condensate and the Polyakov loop fluctuations can be  understood
semi-analytically.
{}From Eqs. (\ref{LLDirac})-(\ref{LabsDirac}), it is clear, that
the contribution of the low-lying Dirac-modes with
$|\lambda_n|\simeq0$ is suppressed,
relative to the higher-lying Dirac modes,  due to the damping factor
$\lambda_n^{N_\tau-1}$.
In fact,
a Dirac matrix element $\langle n|\hat{U}_4|n \rangle$
does not yield a stronger singularity than $1/\lambda_n^{N_\tau-1}$,
therefore the contribution from the low-lying Dirac modes to the Polyakov loop \cite{DSI}, as well as, to its fluctuations is negligible.
Hence,
the essential modes for chiral symmetry breaking do not contribute
to a sensitive probe for   deconfinement  in QCD.
Thus, this finding suggests 
no direct one-to-one correspondence between confinement and
chiral symmetry breaking in QCD.

\section{summary and conclusions}
The main objective of these studies was to establish  the relation between
the Polyakov loop and its fluctuations with the eigenmodes of a Dirac operator.
Based on the lattice QCD formalism, we have derived a Dirac spectral representation
of the real, imaginary and the modulus of the Polyakov loop and their fluctuations.
Although the formulation was done on a temporally odd-number lattice, this choice
of the parity for the lattice size does not alter the physics in the continuum
limit with any large number of $N_\tau$.
%
%
The analytical decomposition of the Polyakov loop and its fluctuations 
is fully general. It is independent from the gauge group, the implementation of
fermions on the lattice, and is also valid at finite baryon density.

To quantify the influence of Dirac modes over the Polyakov loop fluctuations,
we have performed Monte Carlo simulations in the SU(3) lattice QCD.
Our calculations   were  carried out
with the standard plaquette action at the quenched level on
$(10^3\times5)$--size lattice at two different  temperatures, corresponding to
confined and deconfined phases.

We have shown that the low-lying Dirac modes have negligible contribution
to the Polyakov loop fluctuations. This result is intact both in confined and
deconfined phases.
On the other  hand, the low-lying Dirac modes are essential, in both phases,
to quantify the chiral condensate.

These findings, 
both in analytical formulas and in numerical calculations, 
suggest 
no direct, one-to-one correspondence between confinement 
and chiral symmetry breaking in QCD.
However, this does not exclude a coincidence of these two properties in QCD 
since the abrupt change of the ground state from chiral symmetry broken 
to restored phase may drive the onset of deconfinement.

The above conclusion is based on the numerical simulations 
on a rather small-size lattice, 
being far from a continuum limit.
Thus, our result on the Polyakov loop fluctuations suffers from 
finite size effects. 
Such effects can certainly modify the values of 
fluctuations at a given temperature, however will not 
change our conclusion on the influence of the low-lying Dirac modes 
on their properties.
The low-lying Dirac modes have negligible contribution 
to the Polyakov-loop fluctuations 
because of the damping factor $\lambda_n^{N_\tau-1}$
which appears 
in Eq.(\ref{RelOrig}). 
Although our numerical calculation was performed at the quenched level, 
the derived analytic formulae are fulfilled 
even in the presence of dynamical quarks. 
It is one of the future prospects to perform full-QCD simulations in the 
present formalism 
to further justify our conclusion. 

In addition,
the derived analytic relations connecting the Polyakov loop 
and Dirac modes are mathematically exact 
for arbitrary odd temporal size $N_\tau$.
Thus we expect that our conclusion is rather robust in the continuum limit \cite{SDI,DSI}. 
Moreover, the ambiguity of the multiplicative renormalization 
of the Polyakov loop has been avoided 
in the ratio of the Polyakov loop susceptibilities. 
Yet, it is left as an important but difficult task to extrapolate 
these analytic relations to the continuum. 

In addition to the Polyakov loop fluctuations, there are further observables
which are linked to deconfinement properties in QCD and show abrupt, but smooth
change across the chiral crossover. One of such observables is the kurtosis of
the net-quark number fluctuations~\cite{EKR06,CSchmidt12,CSasaki14}. Besides,
the QCD monopole, in the maximally Abelian gauge,  is
a relevant degree of freedom in the low-energy QCD~\cite{Miyamura,Woloshyn}, and
plays a fundamental role for non-perturbative phenomena
such as confinement and chiral symmetry breaking.
Thus, from the future perspectives, it would be of particular interest to investigate
such quantities in terms of the Dirac-mode expansion and to explore the influence
of the low lying eigenmodes on their properties  near the  chiral crossover.

\section*{Acknowledgements}
K. R. acknowledges partial support of the U.S. 
Department of Energy under Grant No. DE-FG02-05ER41367, 
and fruitful discussions with Bengt Friman and Pok Man Lo. 
T.M. D. is supported by Grant-in-Aid for 
JSPS Fellows (Grant No. 15J02108), and H. S. is supported 
by the Grants-in-Aid for Scientific Research (Grant 
No. 15K05076) from Japan Society for the Promotion of 
Science. The work of K. R. and C. S. has been partly 
supported by the Polish Science Foundation (NCN) under 
Maestro Grant No. DEC-2013/10/A/ST2/00106 and by the 
Hessian LOEWE initiative through the Helmholtz 
International Center for FAIR. The lattice QCD calcula- 
tions were performed on NEC-SX8R and NEC-SX9 at Osaka University.

\end{document}